\documentclass[12pt]{article}

\pdfoutput=1

\usepackage{graphics}
\usepackage{epsfig}
\usepackage{amsfonts}
\usepackage{amssymb}
\usepackage{amsmath}
\usepackage{dsfont}
\usepackage{pifont}
\usepackage{bbm}
\usepackage{multirow}
\usepackage{latexsym}
\usepackage{verbatim}
\usepackage{mcite}
\usepackage{cite}
\usepackage{units}
\usepackage[all]{xy}
\usepackage{cancel}
\usepackage{slashed}
\usepackage{tikz}
\usetikzlibrary{snakes}
\newcommand{\be}{\begin{equation}}
\newcommand{\ee}{\end{equation}}
\newcommand{\ben}{\begin{displaymath}}
\newcommand{\een}{\end{displaymath}}
\newcommand{\bea}{\begin{eqnarray}}
\newcommand{\eea}{\end{eqnarray}}

\newcommand{\bean}{\begin{eqnarray*}}
\newcommand{\eean}{\end{eqnarray*}}

\DeclareMathAlphabet{\mathpzc}{OT1}{pzc}{m}{it}

\begin{document}
\pagestyle{plain}

%----------------------------------------------------------------------%
%  numbering sections, equations, footnotes, etc...
%----------------------------------------------------------------------%

\makeatletter \@addtoreset{equation}{section} \makeatother
\renewcommand{\thesection}{\arabic{section}}
\renewcommand{\theequation}{\thesection.\arabic{equation}}
\renewcommand{\thefootnote}{\arabic{footnote}}

%----------------------------------------------------------------------%
%  Resetting of counters
%----------------------------------------------------------------------%

\setcounter{page}{1} \setcounter{footnote}{0}

%----------------------------------------------------------------------%
%  title page
%----------------------------------------------------------------------%

\begin{titlepage}

\begin{flushright}
UUITP-39/18\\
\end{flushright}

\bigskip

\begin{center}

\vskip 0cm

{\LARGE \bf The quantum swampland} \\[6mm]

\vskip 0.5cm

{\bf Ulf Danielsson\,  }\let\thefootnote\relax\footnote{{\tt ulf.danielsson@physics.uu.se}}\\

\vskip 25pt

{Institutionen f\"or fysik och astronomi, University of Uppsala, \\ Box 803, SE-751 08 Uppsala, Sweden \\[2mm]}

\vskip 0.8cm

\end{center}

\vskip 1cm

\begin{center}

{\bf ABSTRACT}\\[3ex]

\begin{minipage}{13cm}
\small

In this paper we propose a quantum version of the swampland conjecture. We argue that quantum instabilities of de Sitter space discovered using field theoretical methods, are directly related to the difficulties in finding stringy de Sitter vacua. 
\end{minipage}

\end{center}

\vfill

\end{titlepage}

%%%%%%%%%%%%%%%%%%%%%%%%%%%%%%%%%%%%%%%%%%%%%%%%%%%%%%%%%
%%
%%               Contents
%%
%%%%%%%%%%%%%%%%%%%%%%%%%%%%%%%%%%%%%%%%%%%%%%%%%%%%%%%%%

\tableofcontents

\section{Introduction}
\label{sec:introduction}

Recently it has been conjectured that there are no metastable dS in string theory, \cite{Danielsson:2018ztv},\cite{Obied:2018sgi}. In \cite{Obied:2018sgi} a specific constraint was conjectured stating that any effective field theory potential derived from string theory must obey
\be
cV \leq \mid \nabla V\mid , \label{SC}
\ee
where derivatives refer to the moduli of the theory, and $c$ is a positive constant of ${\cal O} (1)$. The constraint allows critical points at negative values of the potential  but not at positive ones. The main motivation behind the conjecture, is the failure in finding trustworthy dS-vauca, and the notoriuos conspirasies against them uncovered over the years.\footnote{There are also general arguments against the existence of the string landscape, see e.g., \cite{Banks:2012hx}} There are convincing evidence that there are no metastable dS solutions in classical type II string theory where you add at most orientifolds. see, e.g.,\cite{Danielsson:2018ztv}\cite{Timm}\cite{Danielsson:2010bc}\cite{Danielsson:2011au}\cite{Danielsson:2012et}. The rare tachyonic dS that can be found, are likely to be spurious in the sense that if you require quantization of fluxes, you will have to tune, e.g., the string coupling so that quantum corrections will become important, \cite{Danielsson:2011au},\cite{Roupec:2018mbn}. A classical analysis will not do.

Adding non-pertubative quantum corrections to type IIB giving rise to SUSY AdS-vacua, which then can be uplifted to dS using anti-branes, \cite{Kachru:2003aw}, are also questioned. If they are accepted as such, there are possible issues with the stability of the anti-branes \cite{Danielsson:2018ztv}\cite{Bena:2009xk}\cite{Michel:2014lva}\cite{Thomas}
\cite{Blaback:2014tfa}\cite{Danielsson:2014yga}\cite{Danielsson:2016cit}, or possible destabilization of the moduli,\cite{Moritz:2017xto}. Even worse, there are doubts whether the non-perturbative effects can give rise to a SUSY AdS in the first place. According to \cite{Sethi:2017phn}, the semiclassical instantons are not corrections to be integrated out. If the classical background is rolling, as it does, it is argued that more work is needed to take care of them correctly. For a differing point of view, see \cite{Kachru:2018aqn}. 

The proposal that there might be no dS in string theory is hotly debated. See, e.g., \cite{Kachru:2018aqn}\cite{Andriot:2018wzk}\cite{Kallosh:2018nrk}\cite{Andriot:2018ept}\cite{Conlon:2018eyr}\cite{Cicoli:2018kdo}\cite{Heisenberg:2018yae}\cite{Akrami:2018ylq}. Some argue in favour of the conjecture, some against or are at least highly sceptical.  As a way out, \cite{Obied:2018sgi}\cite{Agrawal:2018own} propose that quintessence has to be seriously considered, (see also \cite{Garg:2018reu}, as well as \cite{Kinney:2018nny} for a confrontation with data is the context of inflation). This possibility was investigated already several years ago, \cite{Blaback:2013fca}, in the context of classical type II vacua, but no example with more than a couple of e-foldings worth of accelerated expansion could be found.  If the dS we are looking for is not a fundamental time independent vacuum, the rules of the game could possibly change. This is the motivation behind \cite{Banerjee:2018qey} where our universe is a brane world riding a bubble that is part of a phase transition. It is not clear in what sense the swampland conjecture applies to this setup.

The purpose of this note is to take the conjecture seriously and consider the remaining corner (from a type II point of view), where we need to consider string loop corrections, i.e., quantum corrections in space time. As observed in \cite{Sethi:2017phn}, balancing classical and quantum loop corrections remains a possibility, even though little explored \cite{Berg:2005yu}\cite{Cicoli:2007xp}. For recent considerations in the context of the swampland, see \cite{Dasgupta:2018rtp}. From a physical point of view, one can also argue that it is awkward to ignore quantum loop corrections. The way one usually approaches dS vacua in the string landscape, involves two steps. First, you try to find a classical potential (possibly including non-perturbative contributions) giving rise to a meta-stable dS. The vacua you find should be tunable so that you can obtain a value of the cosmological constant small enough to be compatible with observations. Then you make sure that the loop corrections are so small that they will not affect stability. This could in principle rule out some of the classical/non-perturbative vacua found, but it has still been claimed that a landscape worth of solutions remains. Even though the quantum corrections may be smaller than the pieces of the potential that are important for stability, they could still be large compared to the resulting value of the cosmological constant. In this way, the detailed anthropic fine tuning of the value would depend on the loops, but not the stability itself.

In view of the swampland conjecture it is interesting to investigate whether the loop effects could play a stabilizing role and point a way out of the swampland. As we will argue, they will not. 

\section{A primer on quantum corrections}

Calculating the effective action, the potential generated by quantum corrections is claimed to be of the form, \cite{ColemanWeinberg},
\be
V_{{1-loop}} = \frac{1}{64 \pi^2} \left( \Lambda^4 STr(m^0) \ln \left(\frac{\Lambda^2}{\mu^2}\right) + 2\Lambda^2 STr(m^2)+ STr \left(m^4\ln \left(\frac{m^2}{\Lambda^2}\right) \right) \right) . \label{CW}
\ee
The supertrace sums over all fermonic as well as all bosonic degrees of freedom with a weight $(-1)^{2j} (2j+1)$, where $j$ is the spin. In a spontaneously broken supersymmetric theory, the numbers match so that the first term cancels. As pointed out in \cite{Ferrara:1994kg}, there are even models where supersymmetry is broken in such a gentle way that the second term vanishes as well. While these corrections add to the potential, one would expect them to be expressible in terms of the K{\"a}hler potential $K$ and the superpotential $W$, if supersymmetry is spontaneously broken. The total potential would then still be given by
\be
V=e^K \left( K^{ij}D_i W D_J \bar{W} - 3\mid W \mid ^2 \right) . \label{SG}
\ee
It is well known that the loop corrections to the potential appear only in the K{\"a}hler potential, and thus they are computable from the kinetic term. Such calculations have been performed in, e.g., \cite{Berg:2005yu}, and in \cite{Cicoli:2007xp} it was shown how these corrections nicely leads to corrections to the potential of just the right form. Obviously, the first term in (\ref{CW}) cannot be there if supersymmetry is spontaneously broken, while the rest of the terms are compatible with (\ref{CW}).

Actually, as discovered in \cite{Akhmedov:2002ts}, elaborated in e.g. \cite{Koksma:2011cq}, and extensively reviewed in \cite{Martin:2012bt}, there is more to the story. It turns out that the result crucially depends on what scheme of regularization you use. The simplest possibility is a brutal cutoff in 3-momentum (or energy), which gives an energy density according to
\be
\rho = \frac{1}{2}\int _0 ^\Lambda \frac{dk k^2}{2\pi^2} \sqrt{k^2+m^2}
= \frac{1}{16 \pi^2} \left( \Lambda^4 +\Lambda^2 m^2 \right)+ \frac{1}{64 \pi^2}m^4\ln \left(\frac{m^2}{\Lambda^2}\right) +...  , \label{rho}
\ee
while the pressure becomes
\be
p = \frac{1}{6}\int _0 ^\Lambda \frac{dk k^2}{2\pi ^2} \frac{k^2}{\sqrt{k^2+m^2}}
= \frac{1}{48 \pi^2} \left( \Lambda^4 -\Lambda^2 m^2 \right)- \frac{1}{64 \pi^2}m^4 \ln \left(\frac{m^2}{\Lambda^2}\right) +... \label{p}
\ee
In this way of defining things neither the first, nor the second term, obey $p=-\rho$, and therefore break the symmetries of dS space. The calculation cannot be argued to support the existence of more than the term proportional to $m^4$, while the presence of the other terms can just be an artifact of the non-covariant cutoff. Support for this interpretation comes from dimensional regularization, where only the dS-invariant third term survives. 

If we instead use a covariant cutoff in 4-momentum, we can make contact with (\ref{CW}). If we take a derivative with respect to $m^2$, we find
\be
\frac{d \rho}{d m^2} =\frac{1}{4}\int \frac{d^3 k}{(2\pi)^3 \sqrt{k^2+m^2}}=\frac{1}{2}\int \frac{d^4 k}{(2\pi )^4}\frac{1}{k^2-m^2+i \epsilon},
\ee
which is manifestly invariant. After a Wick rotation to Euclidean space, we impose a cutoff in the Euclidean momentum $k_E$, integrate back with respect to $m^2$, and obtain
\bea
\rho&=&\frac{1}{2} \int \frac{d^4 k_E}{(2\pi )^4}\ln \left( \frac{k_E^2+m^2}{\mu^2} \right) \\
&=& \frac{1}{64 \pi^2} \left( \Lambda^4 \ln \left(\frac{\Lambda^2}{\mu^2}\right) + 2\Lambda^2 m^2+  m^4 \ln \left(\frac{m^2}{\Lambda^2} \right) \right) +...,
\eea
where we have only kept the leading divergences. We have added an $m$ independent constant of integration (parametrised by the arbitrary scale $\mu$) to match the results of \cite{ColemanWeinberg} for a massive scalar without interactions. (In a path integral approach to the effective potential, the dimensionful scale $\mu$ is needed when defining the measure). Can we trust these new terms? The first (the quartic one) is really not determined at all, since it appears as a constant of integration and does not contribute with any field dependence to the effective potential. The status of the second is less clear.

If we, following \cite{Akhmedov:2002ts}, turn back to the expressions for $\rho$ and $p$ with a cutoff in energy, we note that the integrands obey $\left( 1-2\frac{d}{d m^2} \right) \rho = 3p$. Assuming $p=-\rho$, we find $m^2 \frac{d \rho}{d m^2} = 2 \rho$ implying that $\rho$ must go like $m^4$. It is the presence of the cutoff in energy that induces the terms with a different equation of state, inconsistent with a pure cosmological constant. Energy is not invariant, and in particular it redshifts in an expanding universe. In the calculation leading to the effective action, the cutoff $\Lambda$ is an invariant mass rather than an energy, and it is thus not unreasonable that the cutoff can yield terms contributing to a cosmological constant.

There is no real agreement in the literature on these important issues.  \cite{Martin:2012bt}, for instance, argues that it is only the $m^4$ term that makes sense. The stringy calculations in e.g. \cite{Berg:2005yu}\cite{Cicoli:2007xp} would support the results obtained from the effective potential with the invariant cutoff given by the Kaluza-Klein masses. Some, as \cite{Marsh:2018kub}, choose for simplicity to focus on the $m^4$ terms since there are models, as noticed in \cite{Ferrara:1994kg}, where only these are present.

Nevertheless, there is an important, but neglected, point connected with the choice of vacuum. The calculation leading to (\ref{CW}) is based on an Euclidean continuation which presupposes a particular choice. We will now turn to these subtleties.

\section{The importance of vacuum choice}

The choice of quantum vacuum does not play an important role in the search of stringy vacua. This is understandable. String field theory is not well developed, and the focus is on effective potentials where quantum corrections may or may not appear. Nevertheless, there are developments within field theory with bearing on the swampland that needs to be considered.

There are several interesting issues that arise doing quantum field theory in de Sitter space. First, the vacuum is not unique. Contrary to Minkowski space, where there is a unique vacuum respecting the symmetries of the background, there is a whole family of dS-vacua, sometimes called the {\it $\alpha$-vacua}, \cite{CT}\cite{alpha} (for a review, see \cite{Danielsson:2002qh}). The issue of choice arises already in inflation. There, one usually picks what is often called the {\it Bunch-Davies vacuum}, \cite{CT}\cite{BD}. The argument goes like this. Expand, e.g., the inflaton in momentum modes, and focus on a particular one of them. Trace it back in time until its wavelength is so short and blueshifted compared to the Hubble scale that you can ignore the fact that you are in dS. Then there is a unique vacuum, the Minkowski one. Pick it, and repeat the construction for all momentum modes. In this way you find the Bunch-Davies vacuum. The crucial catch, important for our discussion of the quantum swampland, is that this naively favored vacuum is likely to be the wrong choice.

In \cite{Danielsson:2002qh}\cite{Danielsson:2002kx}\cite{Danielsson:2002mb} it was argued that there are good reasons to pick another vacuum than the Bunch-Davies\footnote{For related ideas in the context of modified CMB-spectrum, see \cite{Martin:2000xs}\cite{Easther:2001fi}\cite{Brandenberger:2002nq}\cite{Brandenberger:2004kx}.}, while  \cite{Polyakov}\cite{Mottola1}\cite{Mottola} have argued that the Bunch-Davies vacuum is fundamentally inconsistent and cannot be sustained. See also \cite{Brandenberger:2018fdd} as well as \cite{Markkanen:2017abw} and \cite{Brahma:2018hrd} for related work. To see why, we use an alternative definition of the Bunch-Davies vacuum obtained through continuation from Euclidean dS, which is just a sphere. If you define a Hamiltonian using the time coordinate of the static patch, you find a thermal spectrum with Hawking temperature  of the order the Hubble constant $H$. This is the analogue of a black hole in equilibrium with a heat bath at the Hawking temperature. In reality, black holes will radiate into empty space and shrink in size with time. Similarly, as argued in \cite{Polyakov}, dS could be sustained if there was a mirror at the cosmological horizon reflecting back all radiation. Without such an artificial setup the Bunch-Davies vacuum obtained using Euclidean dS is simply not the correct vacuum. \cite{Polyakov}\cite{Mottola1}\cite{Mottola} argue that the analysis has to be redone and that there will be particle production with a decay of the cosmological constant. The rate is difficult to compute but it is non-zero and might be fast. 

The intuitively best way to understand what is going on is perhaps given in \cite{Mottola}. The relevant comparison is with the Schwinger effect in a constant electric field. Through tunneling pairs of electrons and positrons are created moving off towards infinity eventually discharging the whole system. \cite{Mottola} concludes that a similar phenomenon will take place in dS with a decay of the cosmological constant, rather than the electric field, as a result. You see why when you realize that there is a puzzle already in the case of an electric field. The starting point is fully time reversal invariant amd there is a priori no reason why a particular direction in time would be selected so that the system could decay. Nevertheless, noone doubts this is what will happen. Methods based on canonical quantization \cite{canonical}, showed that the result follows using the same $m^2 \rightarrow m^2 - i\epsilon$ prescription as is behind the causal propagator of Feynman. This is what selects a particular direction in time and allows for a decay. The point of \cite{Mottola} is that real time quantization and Euclidean continuation are physically inequivalent.  The exact same reasoning applies to dS. If you start out with Euclidean dS, and analytically continue to Lorentz signature, you obtain the Bunch-Davies vacuum and no decay. As demonstrated in \cite{Mottola}, the Bunch-Davies vacuum is contrived and unphysical and barely deserves to be called a vacuum. It is more like a fine tuned superposition of particle and anti-particle modes so that particle creation is compensated for by particle annihilation. Particles are shot in from infinity so that they exactly catch and annihilate the particles being created. This is not what you expect from a set of natural initial conditions. One should also keep in mind that from the point of view of cosmology, it is not global dS but rather the Poincare patch that is the more relevant one.

\cite{Danielsson:2002kx} gave a complementary way of arguing against Bunch-Davies based on the the physical motivation for this particular vacuum. If there is a fundamental cutoff in energy, $\Lambda$, you cannot really trace a given mode any further back than to when its wavelength is as short as the fundamental scale, be it Planck, string or something else. At that scale, the argument for which vacuum to choose breaks down and you need to find another principle. The simplest choice, as advocated in \cite{Danielsson:2002kx}, would be to assume the {\it instantaneous Minkowski vacuum} for the inflaton or any other scalar field. This is defined by
\be
a_{k}\left( \eta _{k,0}\right) \left| 0,\eta _{k,0}\right\rangle =0.
\ee
Here we have defined
\be
\mu _{k}\left( \eta \right)  =f_{k}\left( \eta \right) a_{k}\left( \eta
_{0}\right) +f_{k}^{\ast }\left( \eta \right) a_{-k}^{\dagger }\left( \eta
_{0}\right) ,
\ee
where $\phi=\mu/a$ is the massless scalar field. We work in conformal time defined by $\eta =-\frac{1}{aH}$ so that $\mu_k$ solves
\be
\mu _{k}^{\prime \prime }+\left( k^{2}-\frac{a^{\prime \prime }}{a}\right)
\mu _{k}=0 
\ee 
The moment when a given mode emerges at the fundamental scale is given by $\eta_{k,0}=-\frac{\Lambda}{Hk}$. Note that this is not an initial condition imposed at a given moment in time for all states, but a condition continously at work at all times as mode after mode appears. When $\Lambda \rightarrow \infty$, and $\eta_{k,0} \rightarrow -\infty$, one recovers the Bunch-Davies vacuum. Each mode turns out ot be a non-trival Bogolubov transformation with a mixing between the creation and annihilation operators of order $H/\Lambda$. Formally, this vacuum might look like a member of the family of dS-invariant $\alpha$-vacua thanks to the way the initial condition of a particular mode is imposed. This is of course strictly speaking not true. First, it is not meaningful to talk about a vacuum above the cutoff, second, the back reaction will cause a decay of the cosmological constant and thus break the symmetries of dS and make $H$ time dependent.  

The dependence on choice of vacuum that we have described, is an example of the UV-sensitivity and non-renormalizability of gravity. What happens at the highest energy scales cannot easily be decoupled from low energies.

It is not difficult to estimate the back reaction in a simple toymodel. The main idea is based on the observation that the cosmological constant has a special status in the sense that it can be viewed as a constant of integration. It can also be moved between the two sides of the Einstein equation. The key is to choose the right pair of linear combinations out of the three linear dependent Friedmann equations as a starting point. Followng the thermodynamical approach to gravity outlined in \cite{Jacobson:1995ab}, applied to a cosmological setting, it is natural to choose, \cite{transplanck}
\bea
\dot{H}=-\frac{4\pi }{M_{p}^{2}}\left( \rho +p\right) , \\ \label{thermoF}
\overset{\cdot }{\rho }+3H\left( \rho +p\right) =0  .
\eea
The second equation is just the continuity equation, while the first tells you how the horizon grows when you throw matter towards it. It is nothing else than the entropy-area relationship, i.e., $\frac{dQ}{dt}=T\frac{dS}{dt}=A\left( \rho +p\right)$, where $S=\frac{M_{p}^{2}}{4}A$ and $T=\frac{H}{2\pi }$ is the dS temperature. The argument does not use this correspondence any further, but may serve as a motivation for why these equations is the preferred choice. The crucial point is that nowhere in these equations can you find the cosmological constant explicitly. It is only when the first equation is integrated, using the second, that the cosmological constant appears as a constant of integration. This hints that it plays a different role than other matter contributions ,and that it is misleading to simply refer to it as dark energy.

The choice of a non-standard vacuum, such that the one suggested in \cite{Danielsson:2002kx}, will lead to particle production. This was discussed in \cite{Starobinsky:2002rp}, where limits from observations were confronted. If there is particle creation there is the issue of energy conservation. This will be guaranteed by the Einstein equations through self consistency, as is especially clear with the choice of equations that we just made. The contribution to the vacuum energy from a non-standard vaccum will be on top of the contributions that we discussed earlier. Focusing on a massless field, where we initially have no contribution, the extra energy density will be given by 
\be
\rho _{\Lambda }= \frac{1}{4\pi ^{2}}\int_{0}^{\Lambda }dkk^{3}\frac{H^{2}}{\Lambda ^{2}}=\frac{\Lambda ^{2}H^{2}}{16\pi ^{2}}  .
\ee
What one needs to keep in mind is that as the universe expands, modes are redshifted downwards, while new ones need to be fed into the system at the cutoff. Let us in the following couple of equations denote the physical momentum with $p$. From
\be
\rho _{\Lambda }\left( a\right) =\frac{1}{4\pi ^{2}}\int_{\varepsilon
}^{\Lambda }dpp^{3}\frac{H^{2}\left( \frac{ap}{\Lambda }\right) }{\Lambda
^{2}}=\frac{1}{4\pi ^{2}}\frac{\Lambda ^{2}}{a^{4}}%
\int_{a_{i}}^{a}dxx^{3}H^{2}\left( x\right) ,
\ee
it follows that the continuum equation aquires a source and becomes
\begin{equation}
\dot{\rho}_{\Lambda }+4H\rho _{\Lambda }=\frac{1}{4\pi ^{2}}\Lambda
^{2}H^{3}. \label{eq:contQ}
\end{equation}
The interpretation of the lower integration limits, is that we consider radiation created later than at a scale factor $a_i$. The modes created at $a_i$ has an energy that is redshifted to $\varepsilon$ at scalefactor $a$. The dependence on the scale factor $a$ is compatible with the equation of state suggested by (\ref{rho}) and (\ref{p}). This is the only place where we need to modify anything in our equations. (\ref{thermoF}) stays the same given its thermodynamic interpretation. Feeding this into the Friedmann equations yield
\be
H^{2}=C_{1}a^{-2n_{1}}+C_{2}a^{-2n_{2}} ,
\ee
where
\be
n_{1,2}=1\pm \sqrt{1-\frac{2\Lambda ^{2}}{3\pi M_{p}^{2}}}.
\ee
Note that there are two constants of integration. The first one simply tells how much radiation there is, while the second is an analogue to the cosmological constant,
We note that the radiation is decaying a bit slower than $1/a^4$ due to the particle creation, and the cosmologcial constant is decaying due to drainage.

\section{The quantum swampland}

We have seen that there is a rather universal problem with dS at the quantum level. The Bunch-Davies vacuum turns out to be unphysical, and another vacuum neds to be selected. The exact one may depend on fine details through UV-sensitivity, but we have at least been able to propose a candidate parametrized by a fundamental scale. If we move out of the classical swampland, in search for a metastable dS where quantum contributions play a role, we have to face this problem. In fact, even if it had been possible to find a vacuum already at the classical level, the quantum contributions are still expected to be much larger than our fine tuned value of the cosmological constant.

The field theoretical difficulties we have outlined, suggest that any calculation in string theory leading up to a result in line with (\ref{CW}) using a covariant cutoff based on Euclidean continuation, need to be revisited. If dS is unstable, there is a spontaneous breakdown of dS invariance at the quantum level similar to the onset of the Schwinger effect in the context of a constant electric field. The resulting particle production will through conservation of energy drain the cosmological constant and induce a time dependence. There does not seem to be a way around this fact. In this way a preferred frame appears, and a cutoff in 3-momentum will do. From the point of view of field theory, this can self consistently undermine the reasoning leading to the presence of a constant dark energy. In a time dependent background, there is no reason to exclude terms with $p \neq - \rho$ based on lack of the right symmetry.  In fact, consistency of the field equations will require them to be present.

Once the vacuum energy starts to change due to quantum drainage, or cloaking of the effective cosmological constant, the balance with respect to the classical moduli will be perturbed. When this happens, there will be an onset of a classical rolling as well. The process can not end until the cosmological constant has reached zero, or below, together with a possible decompactification of the extra dimensions. As the Hubble constant decreases in value, the quantum effects abate.

We conclude that the inconsistency of the Bunch-Davies vacuum supports a quantum version of the swampland conjecture. The speed of the decay is sensitive to the details of the model, and a careful -- possibly difficult --analysis is needed to determine how important the effect is. In particular, one must study string theory in time dependent backgrounds in the presence of non-trivial vacuum states. This is in line with the arguments put forward in \cite{Sethi:2017phn}.

It is amusing to attempt to make a direct connection with the swampland conjecture as formulated in \cite{Obied:2018sgi}. While the quantum decay cannot necessarily be translated into a slope of a potential, one can still make the comparison using the slow roll relation
\be
-\frac{\dot{H}}{H^2} = \frac{M_p^2}{2} \left( \frac{V'}{V} \right)^2 .
\ee
From this it immediately follows that $c$ as defined through (\ref{SC}) is given by
\be
c=\sqrt{\frac{2}{3}}\frac{\Lambda}{M_p} ,
\ee
when $\Lambda$ is much smaller than $M_p$.
In a particular model, the actual value of $c$ will be dominated by either classical or quantum contributions, but the conjecture would be that there is a minimal value.\footnote{These results are similar to what is claimed in \cite{Dvali:2018fqu}.}

It is possible that the effect that we have studied, can be mapped in interesting ways through the string dualities. It could equally well affect the non-perturbative terms as the loop ones. While dualities are best trusted when in the presence of supersymmetry, one could still obtain valuable clues. Perhaps stringy dualities together with the conceptual insights of why the Bunch-Davies vacuum cannot be used, can pave the way for a proof of the swampland conjecture or suggest ways around it.

\section{Conclusion}

We have briefly reviewed arguments for why there are no classical dS, as well as no compelling examples of dS even if you add non-perturbative effetcs. The only loop hole would be the addition of qauntum loop effects. If true, this means that you would need to move out of the corner of parameter space where these effects can be ignored, develop tools to calculate them, and try to carefully balance them against the classical terms in order to obtain a metastable dS. The goal of this paper has been to point to results from field theory that already suggest that this project will fail. The only vacuum that is compatible with time independent, stable dS, is the Bunch-Davies vacuum that is argued to be unphysical. Using general arguments, particle production is inevitable and the contribution to the dark energy due to quantum loops will be drained. This reasoning suggests that the cosmological constant is cloaked through a time dependent renormalization and will approach zero. To be explicit, we have estimated the qunatum contribution to the constant $c$ iin a simplified setting. 

Given the arguments put forward in \cite{Polyakov:2012uc}\cite{Mottola1}\cite{Mottola}, it seems hard to find a way around these difficulties. The connection with the concerns raised in, e.g., \cite{Sethi:2017phn} is tantalazing.

We hope that these ideas can serve as an inspiration to explore the quantum swampland and beyond.

\section*{Acknowledgments}

We would like to thank Paul Anderson, Giuseppe Dibitetto, Fawad Hassan, Emil Mottola, and Thomas van Riet for discussions. The work of the author was supported by the Swedish Research Council (VR).

\end{document}